\documentclass[aps,10pt,twocolumn,showpacs,prl,floatfix]{revtex4-1}
\usepackage{graphicx}
\usepackage{amssymb,amsmath,tabularx}
\usepackage{bm}

\begin{document}

\title{Stochastic predator-prey dynamics of transposons in the human genome}

\author{Chi Xue}
\affiliation{Department of Physics, Center for the Physics of Living Cells and Institute for Universal Biology, University of Illinois at Urbana-Champaign \\
Loomis Laboratory of Physics, 1110 West Green Street, Urbana, Illinois 61801-3080, USA}
\affiliation{Carl R. Woese Institute for Genomic Biology, University of Illinois at Urbana-Champaign \\
1206 West Gregory Drive, Urbana, Illinois 61801, USA }

\author{ Nigel Goldenfeld}
\affiliation{Department of Physics, Center for the Physics of Living Cells and Institute for Universal Biology, University of Illinois at Urbana-Champaign \\
    Loomis Laboratory of Physics, 1110 West Green Street, Urbana, Illinois 61801-3080, USA}
\affiliation{Carl R. Woese Institute for Genomic Biology, University of Illinois at Urbana-Champaign \\
    1206 West Gregory Drive, Urbana, Illinois 61801, USA }

\date{\today}

\begin{abstract}
Transposable elements, or transposons, are DNA sequences that can jump from
site to
site in the genome during the life cycle of a cell, usually encoding
the very enzymes which perform their excision. However, some transposons are
parasitic, relying on the enzymes produced by the regular transposons.  In this
case, we show that a stochastic model, which takes into account the
small copy numbers of the transposons in a cell, predicts noise-induced
predator-prey oscillations with a characteristic time scale that is
much longer than the cell replication time, indicating that the state
of the predator-prey oscillator is stored in the genome and transmitted
to successive generations. Our work demonstrates the important role of number
fluctuations in the expression of mobile genetic elements, and shows
explicitly how ecological concepts can be applied to the dynamics and
fluctuations of living genomes.
\end{abstract}

\pacs{87.23.Cc, 87.10.Mn, 87.23.Kg}

%05.40.-a   Fluctuation phenomena, random processes, noise, and Brownian motion
%87.10.Mn   Stochastic modeling
%87.23.Kg   Dynamics of evolution
%87.23.Cc   Population dynamics and ecological pattern formation

\maketitle

Transposable elements (TE) \cite{mcclintock1950origin,
mcclintock1953induction} or transposons are DNA sequences that can
migrate from site to site in a host genome. These mobile genetic
elements are found in all three domains of life, but especially compose
a significant fraction of eukaryotic genomes, for example, occupying
45\% of the human genomic sequence \cite{Initial_Human_2001}.
Transposons are regarded as a major driver of adaptation and evolution
\cite{kazazian2004TEdrive_evolution}, since they can induce both
beneficial and deleterious transformations in the host genome, by
inserting into encoding or regulation sequences, or causing misaligned
pairing and unequal crossovers of chromosomes. In most cases, the
modifications are disadvantageous to the host, for example causing
hemophilia A in humans \cite{dombroski1991isolation_L1}.
The activity of TEs has historically been observed through detailed
population level assays, but recent measurements have demonstrated
their activity in real time in living cells, using sophisticated
fluorescence techniques \cite{Kim28062016}, quantifying in detail how
the stochastic processes of excision are not purely random, but reflect
a cell's environment and genetic history. The interplay between
transposon dynamics and replication, the cell's genotype and phenotype
and the interactions with the environment are all reminiscent of
population dynamics of organisms within an ecosystem, and this
perspective is one that we explore and quantify here.

The dynamics of TEs are complex, but can be conveniently separated into
two types of edit operation on the host genome: copy-and-paste, and
cut-and-paste \cite{wicker2007classification}. DNA transposons cut
themselves out of the original site on the genome and later reintegrate
at another site, thus performing a \lq\lq cut-and-paste" operation
which leaves the genome size invariant. Retrotransposons first
transcribe into mRNA intermediates and then retrotranscribe to a new
site on the genome sequence. This \lq\lq copy-and-paste" dynamics leads
to the growth of the genome size.  Some transposons (autonomous) encode
the very enzymes which perform their excision, while others are
parasitic (non-autonomous), relying on the enzymes produced by the regular TEs.

Several theoretical approaches have been proposed to study the dynamics
of transposons. Population genetics models
\cite{charlesworth1983populationTE, charlesworth1994populationTE,
langley1983transposable_theory, brookfield1997populationTE,
leRouzic2006populationTE, le2005models_popu_TE} were first developed to
describe the equilibrium distribution of transposons in a population.
Recent development views the genome as an ecosystem, with genetic
elements of different types playing the role of individuals from
different species
\cite{brookfield2005ecologyTE,venner2009TE_ecology,le2007genome_eco,
le2007genome_eco,leRouzic2007long_term_TE,serra2013neutral,linquist2015applying}.
In the case of non-autonomous transposons, a mean-field predator-prey
type model describes their parasitic relationship with an autonomous
transposon \cite{le2007genome_eco,leRouzic2007long_term_TE}.
However, these models do not account for the molecular level
interactions between transposable elements and the dynamic behavior
turns out to be sensitively dependent on these details. Furthermore, in
a cell the copy number fluctuations are large, since the number of
active (expressed) transposons is usually of order ten to a hundred
\cite{Kazazian_2003_HotL1}.  Thus, the next generation of transposon
models needs to take into account molecular details and stochasticity.

The purpose of this Letter is to develop a minimal individual-level
model based on the specific interaction mechanism between a pair of
autonomous-non-autonomous transposons.  We begin with a model of the
interactions between the TEs, and then use techniques from statistical
mechanics to derive stochastic differential equations 
\cite{mckane2005predator-prey}. Our model predicts that number
fluctuations generate persistent, noisy oscillations
 in the populations
of the TEs, with a characteristic time scale that is much longer than
the cell replication time, indicating that the state of the
predator-prey oscillator is stored in the genome and transmitted to
successive generations.  Our work builds upon recent results that have
shown how demographic stochasticity in ecosystems, where population
size is integer-valued and locally finite, can lead to minimal models
of persistent population cycles \cite{mckane2005predator-prey} or
spatial
patterns \cite{butler2009robust,biancalani2010stochastic,tauber2012population, 
houchmandzadeh2014remarkable, fort2013statistical} 
without extra assumptions about the details of predation.

\begin{figure}
    \includegraphics[width = 1\columnwidth]{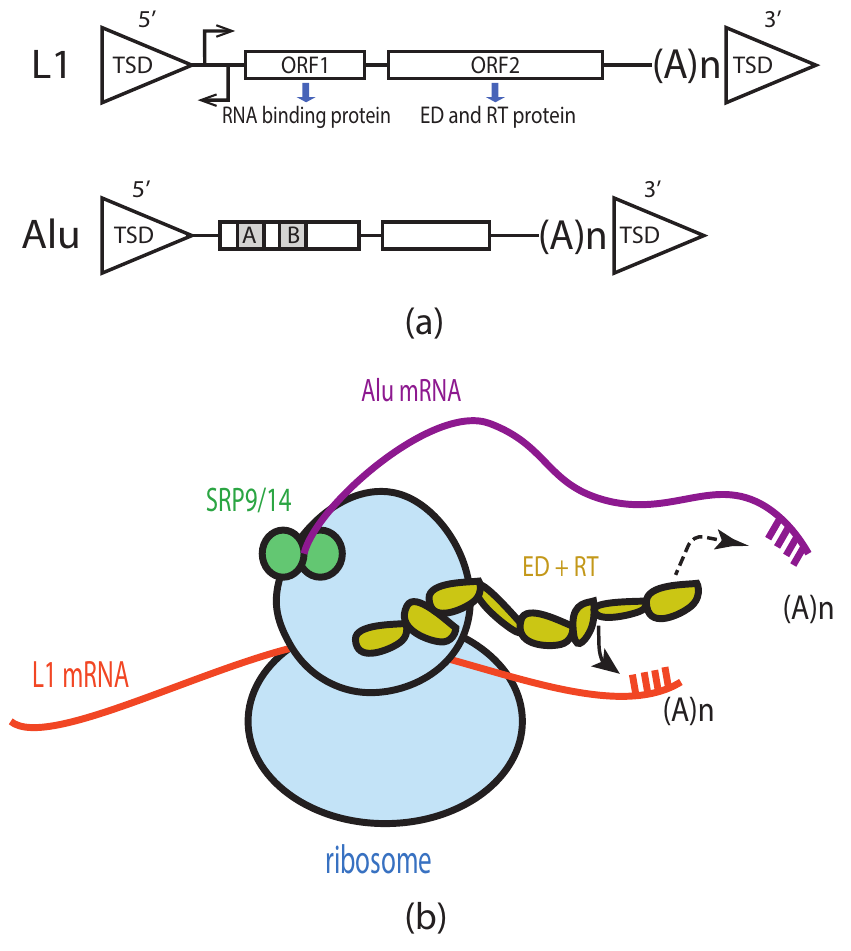}
    \caption{(Color online.) (a) shows the structure of L1 and Alu elements. L1 has a pol
        II promoter (the right-pointing arrow) and an antisense promoter (the
        left-pointing arrow) followed by two open reading frames (ORFs) encoding a RNA binding
        protein and a protein that consists of an endonuclease (ED) and a
        reverse transcriptase (RT), respectively. Alu is composed of two
        non-coding monomers, with the left one bearing A and B boxes (the
        shaded area in the figure)  as the pol III promoter. L1 and Alu
        elements share similar poly-A tails and are both flanked by target site
        duplicates (TSDs). (b) shows the \textit{cis} and \textit{trans} effects
        of L1 elements. When an ED+RT protein is translated at the ribosome, it
        \textit{cis}-preferentially binds with the L1 mRNA that codes it,
        indicated by the solid arrow. An Alu mRNA can bind with two signal
        recognition proteins SRP9 and SPR14, and then attach to the ribosome.
        The nascent ED+RT protein then can \textit{trans}-bind to the Alu mRNA,
        which has a similar poly-A tail (indicated by the dashed arrow),
        presumably with a similar probability to binding to L1 mRNA.
        }
    \label{fig:LS}
\end{figure}

\smallskip
\noindent \textit{Detailed model for transposon dynamics:-}
Retrotransposons consist of two subgroups: LTR-transposons that have a
long terminal repeat (LTR) structure, and non-LTR transposons that do
not \cite{wicker2007classification}. There are two types of non-LTR
elements that show especially interesting interaction: the autonomous
long interspersed nuclear elements (LINEs), and the non-autonomous
short interspersed nuclear elements (SINEs) \cite{Singer_1982}. In the human genome, the
only active LINEs are LINE1 (L1) elements, which take up 17\% of the
entire genome \cite{Initial_Human_2001}. They help SINEs, such as Alu 
elements, to transpose by providing critical enzymes used in the
copy-and-paste dynamics \cite{mills2007L1_Alu_SVA}.
We take L1 and Alu elements in the human genome as an
example of a LINE-SINE pair and build a model of their interaction.

When a protein is produced at a ribosome coded by an L1 mRNA, it tends
to bind with that particular mRNA, presumably by recognizing its 
polyadenine (poly-A) 
tail \cite{doucet2015L1ployA}, and later retrotranscribes
it into the genome. This is known as the \textit{cis}-preference of
L1 elements \cite{wei2001L1cis_trans}. However, if an Alu mRNA attaches
to the same ribosome, then it can bind with the nascent protein by
faking the L1 mRNA poly-A tail \cite{boeke1997LS_polyA}. In this way,
Alu elements steals the transposition machinery designed by L1 elements
\cite{weiner2002S_L_bite, dewannieux2003line_help_alu}. This is
known as the \textit{trans}-effect of L1 elements
\cite{wei2001L1cis_trans}. The mechanism is sketched in Fig.
\ref{fig:LS}(b).

\smallskip
\noindent \textit{Minimal model for transposon dynamics:-} Based
on the above detailed interaction, an individual level minimal model can be
made, discarding all details about how proteins are made and how complexes
are formed. The individual reactions are shown in Eq. (\ref{reactions}), where $L$ stands
for an active LINE, $S$ for an active SINE, and $R_L$ for the complex
of the ribosome, LINE mRNA and nascent protein. Deactivated transposons
do not participate in the transposition events and thus are excluded
from the model.

An $L$ element encodes the complex $R_L$ at the rate $b_R$. The complex
$R_L$ retro-transposes to produce a new $L$ element at the rate $b_L$,
if there is no interruption. $S$ element hijacks the complex $R_L$ to
duplicate itself at the rate $b_S/V$, where $V$ is the system
size. The complex $R_L$ decays at the rate $d_R$. $L$ and $S$ elements
are deactivated, at the rates $d_L$ and $d_S$, respectively. $\emptyset$
stands for null.  The reactions for this minimal model, with the
corresponding forward rates are as follows:
\begin{subequations} \label{reactions}
\begin{align}
    L & \rightarrow  L + R_L, & &b_R \\
    R_L & \rightarrow  L, & &b_L \\
    R_L + S & \rightarrow  2S, & &b_S/V \\
    R_L & \rightarrow  \emptyset, & &d_R \\
    L & \rightarrow  \emptyset, & &d_L \\
    S & \rightarrow  \emptyset, & &d_S
\end{align}
\end{subequations}
We assume the system is well mixed because mixing of reactants is faster,
happening constantly within the cell lifetime, than the reactions.

\begin{figure}
    \includegraphics[width = 1\columnwidth]{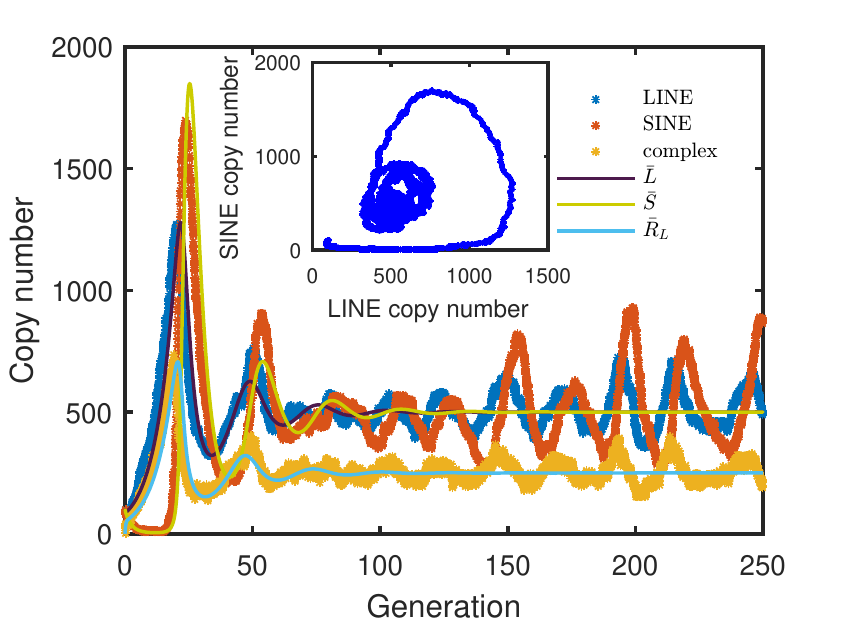}
    \caption{(Color online.) Results of a typical stochastic simulation with illustrative parameters $b_R = 2$, $b_L = 1$, $b_S = 1$, $d_R = 2$, $d_L = 0.5$, $d_S = 0.5$, and the
    system size $V = 500$.
    The main figure shows the copy numbers of active
    LINEs, SINEs and ribosome/L-mRNA/protein complexes as a function of
    time, in the unit of a cell generation. Copy numbers fluctuate around
    constant values, demonstrating quasi-cycles with period $\sim 25$
    generations.
    Solid lines are obtained by evolving the deterministic equations
    and show oscillatory decay toward steady values.
    Demographic noise induces quasi-cycles by constantly stimulating the deterministic oscillation mode.
    The inset shows the trajectory on the $L$-$S$ plane. The
    circular envelope indicates a phase difference of roughly $\pi/2$.
    } \label{fig:popu_LSR}
\end{figure}

We first use the Gillespie algorithm \cite{gillespie1976} to simulate
the above reactions. Copy number vs time curves are plotted in Fig.
\ref{fig:popu_LSR}. As shown in the figure, LINE and SINE copy numbers
fluctuate around constant values, in the form of quasi-cycles. The
circular envelope of the trajectory on the $L$-$S$ plane indicates a
phase difference of roughly $\pi/2$, with SINE lagging LINE,
supporting the identification of SINEs as predators on the LINEs.

\smallskip
\noindent \textit{System size expansion:-}
Let the copy number concentrations of active LINEs, SINEs
and complexes be $L$, $S$ and $R_L$ respectively. Then, with the system
size being $V$, $VL$, $VS$ and $VR_L$ are equal to the actual copy
numbers of the corresponding groups. The master equation about the
probability  $\mathcal{P}(L, S, R_L)$ of the system being in the state
$(L, S, R_L)$ is written down as follows,
\begin{align} \label{master_X}
    & \frac{\mathrm{d}}{\mathrm{d}t}\mathcal{P}(L,S,R_L)  \nonumber  \\
    = & V \Bigl\{ (\mathcal{E}_{R_L}^- - 1) b_R L + (\mathcal{E}_{R_L}^+ \mathcal{E}_L^- - 1) b_L R_L   \nonumber \\
                    & + (\mathcal{E}_{R_L}^+ \mathcal{E}_S^- - 1) b_S R_L S + (\mathcal{E}_{R_L}^+ - 1) d_R R_L       \nonumber \\
                    & + (\mathcal{E}_L^+ - 1) d_L L + (\mathcal{E}_S^+ - 1) d_S S
    \Bigr\} \mathcal{P},
\end{align}
with the raising and lowering operators given by
\begin{equation}\label{Expansion}
\mathcal{E}_X^\pm f(X) \equiv  f(\frac{N_X \pm 1}{V})\approx f(X) \pm \frac{1}{V} \partial_X f + \frac{1}{2V^2} \partial_X^2 f,
\end{equation}
where $f$ is an arbitrary function of the concentration 
$X$, and $X$ stands for $L$, $S$ or $R_L$.

Substituting the expansions of operators into the master Eq.
(\ref{master_X}), and saving terms up to order $O(V^{-1})$, we obtain a
non-linear Fokker-Planck equation.
The corresponding Langevin equations about concentrations $L$, $S$ and $R_L$ 
are nonlinear with multiplicative noises.

To obtain a set of linearized Langevin equations for concentration 
fluctuations, we perform the van Kampen's system size expansion, separating 
concentrations into deterministic part, $\bar{L}$, $\bar{S}$ and $\bar{R}_L$, 
and stochastic part, $\xi$, $\eta$ and $\theta$, as follows.
\begin{equation}\label{vanKampen}
    L = \bar{L} + \frac{\xi}{\sqrt{V}}, \quad S = \bar{S} + \frac{\eta}{\sqrt{V}}, \quad R_L = \bar{R}_L + \frac{\theta}{\sqrt{V}}.
\end{equation}

\begin{figure}
    \centering
    \includegraphics[width = 1\columnwidth]{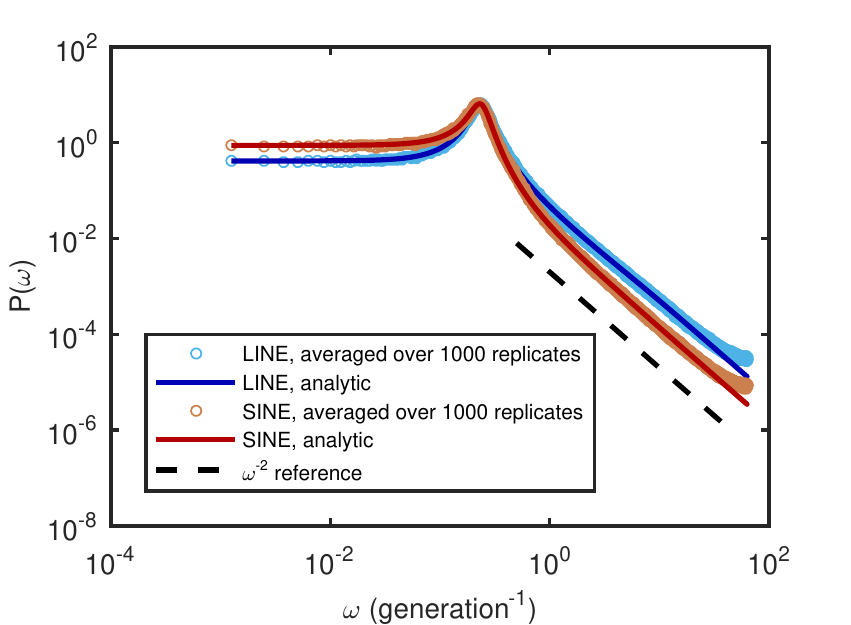}
    \caption{(Color online.) Power spectra of the LINE and SINE         
    concentration fluctuations. Circles stand for the power spectra
    obtained by averaging over 1000 replicates. Solid lines stand for
    the calculated spectra. The dash line is a reference function $\sim
    \omega^{-2}$. Parameters are $ b_R = 2$, $b_L = 1$, $b_S = 1$, $d_R = 2$, $d_L =
    0.5$, $d_S = 0.5$, $V = 500$. The peak angular frequency is equal to $0.23$
    generation$^{-1}$, corresponding to a period of $27$ generations. The
    straight tail, in log-log scale, has a slope of $-2$, indicating a
    $\omega^{-2}$ asymptotic behavior. } \label{fig:spectrum}
\end{figure}

Writing
\begin{equation}
    \Pi(\xi,\eta,\theta) \equiv \mathcal{P}(L,S,R_L),
\end{equation}
we find that
\begin{equation}
    \frac{\mathrm{d}}{\mathrm{d}t}\mathcal{P} = \partial_t \Pi - \sqrt{V} \frac{\mathrm{d}\bar{L}}{\mathrm{d}t} \partial_\xi \Pi - \sqrt{V} \frac{\mathrm{d}\bar{S}}{\mathrm{d}t} \partial_\eta \Pi - \sqrt{V} \frac{\mathrm{d}\bar{R}_L}{\mathrm{d}t} \partial_\theta \Pi.
\end{equation}
Substituting the system size expansion expressions Eq. (\ref{vanKampen}) into the nonlinear
Fokker-Planck equation and matching orders of $V$,
we obtain to $O(\sqrt{V})$
\begin{subequations} \label{deterministic_xyz}
\begin{align}
    \frac{\mathrm{d} \bar{L}}{\mathrm{d}t} &= b_L \bar{R}_L - d_L \bar{L},  \\
    \frac{\mathrm{d} \bar{S}}{\mathrm{d}t} &= b_S \bar{S}\bar{R}_L - d_S \bar{S},  \\
    \frac{\mathrm{d} \bar{R}_L}{\mathrm{d}t} &= b_R \bar{L} - b_L \bar{R}_L - b_S \bar{S} \bar{R}_L - d_R \bar{R}_L.
\end{align}
\end{subequations}
These are the deterministic, or mean field, equations.
The coexistence steady state, where $\bar{L}$, $\bar{S}$ and $\bar{R}_L$ are all non-zero, is always exponentially stable, according to
linear stability analysis. We have verified numerically that the
imaginary part of the linear stability matrix eigenvalues provides a
reasonable estimate for the angular frequency of quasi-cycles.
Specifically, for the parameters used to generate Fig.
\ref{fig:popu_LSR} and Fig. \ref{fig:spectrum}, the eigenvalue
imaginary part is equal to $0.2330$, and agrees well with the Gillespie
simulation value for the peak angular frequency, $0.23$
generation$^{-1}$, of the quasi-cycle power spectra shown in Fig.
\ref{fig:spectrum}.

By matching $O(1)$ terms, we obtain
the linearized Langevin equations for $\xi$, $\eta$ and $\theta$ using Ito's Lemma \cite{ito1944109}:
\begin{subequations} \label{Langevin}
\begin{align}
    \frac{\mathrm{d}\xi}{\mathrm{d}t} &= b_L \theta - d_L \xi + r(t),  \\
    \frac{\mathrm{d}\eta}{\mathrm{d}t} &= b_S \bar{R}_L \eta + b_S \bar{S} \theta - d_S \eta + s(t), \\
    \frac{\mathrm{d}\theta}{\mathrm{d}t} &= b_R \xi - b_L \theta - b_S \bar{R}_L \eta - b_S \bar{S} \theta - d_R \theta + h(t).
\end{align}
\end{subequations}
$r(t)$, $s(t)$ and $h(t)$ are noises in $\xi$, $\eta$ and $\theta$, 
respectively. The correlations between these noises are given by
\begin{subequations} \label{corr_t}
\begin{align}
    \langle h(t)h(t') \rangle &= \delta(t-t')(b_R \bar{L} + b_L \bar{R}_L + b_S \bar{S}\bar{R}_L + d_R \bar{R}_L),    \\
    \langle r(t)r(t') \rangle &= \delta(t-t')(b_L \bar{R}_L + d_L \bar{L}),     \\
    \langle s(t)s(t') \rangle &= \delta(t-t')(b_S \bar{S}\bar{R}_L + d_S \bar{S}),       \\
    \langle h(t)r(t') \rangle &= \delta(t-t')(-b_L \bar{R}_L),         \\
    \langle h(t)s(t') \rangle &= \delta(t-t')(-b_S \bar{S}\bar{R}_L),           \\
    \langle r(t)s(t') \rangle &= 0.
\end{align}
\end{subequations}
These Langevin equations describe the fluctuations of concentrations
 around the steady state values.

\medskip
\noindent \textit{Persistent oscillations:-}
The power spectra
$P_{\xi\xi}(\omega)$ and $P_{\eta\eta}(\omega)$ can be calculated by
manipulating the Fourier transform of Eq. (\ref{Langevin}) and the
correlations Eq. (\ref{corr_t}). The result is a complicated fraction,
of which the numerator is a fourth order polynomial of $\omega$ and the
denominator a sixth order polynomial of $\omega$. Asymptotically, the
power spectra have a tail in the form of $\omega^{-2}$. Figure
\ref{fig:spectrum} shows a comparison between the power spectra
obtained from simulation and the analytic calculation, which demonstrates a satisfactory agreement.
This minimal model shows that the negative feedback of SINEs on LINE
transposition rate results in a predator-prey like dynamics
\cite{mckane2005predator-prey}, with noise induced quasi-cycles.

\smallskip
\noindent {\it Estimation of parameters:-} For the human genome,
transposition rates of L1 and Alu elements measured by the mutation
accumulation method are of order 1 in $O(10) \sim O(100)$ births
\cite{rosenberg2003TE_rates, huang2012activeTE, cordaux2006AluRate}.
The deactivation rates have a lower limit set by the base pair point
mutation rate, which is roughly $10^{-8}$ per base pair per generation
\cite{nachman2000genomic_mutation, roach2010mutation_rate}. These rates
seem to be too slow to generate any experimentally detectable dynamical
behaviors. However, this estimate only accounts for fixed mutations
that are not lethal, and thus underestimates the actual mutation rates.
In a recent experiment \cite{Kim28062016} on real-time transposition
events in living bacterium cells, the actual transposition rate
directly observed was $10^{3}$ times higher than that obtained by the
mutation accumulation method. Moreover, the point mutation rate can be
raised by a factor of $10^{2}$ by deactivating the base pair mismatch
repair machinery \cite{elez2010mutation_real_time}. Thus, for a
single-cell experiment rather than a large population, the relevant
estimate is: $b_R = 2$, $b_L = 1\times 10^{-2}$, $b_S = 1\times
10^{-2}$, $d_R = 1$, $d_L = 1\times 10^{-2}$, $d_S = 1\times 10^{-2}$,
with units being generation$^{-1}$. The resultant quasi-cycle period
should be roughly $1\times 10^{3}$ generations.  Such oscillations
could potentially be observed by integration of
the LINE/SINE elements into a host microbial cell, \textit{E. coli} for example, and
using novel reporter techniques \cite{Kim28062016, Kuhlman2017}.

In conclusion, we have shown that the dynamics of transposons can fruitfully be 
analyzed using analogy to ecological models, equipped with tools from 
statistical physics.
Our calculations predict the existence of potentially observable,
persistent and noisy oscillations in the populations of active SINEs
and LINEs.

\medskip
\begin{acknowledgments}
\noindent\textit{Acknowledgments:-} We acknowledge helpful discussions
with Oleg Simakov and Tom Kuhlman.  This work was partially supported by the National
Science Foundation through grant PHY-1430124 to the NSF Center for the
Physics of Living Cells, and by the National Aeronautics and Space
Administration Astrobiology Institute (NAI) under Cooperative Agreement
No. NNA13AA91A issued through the Science Mission Directorate.
\end{acknowledgments}

\bibliography{SINE_LINE_ref_v2}
\bibliographystyle{apsrev4-1}

\end{document}